\documentclass[12pt]{article} 

\usepackage{geometry}    
\usepackage{titlesec}    
\usepackage{setspace}    
\usepackage{authblk}     
\usepackage{lipsum}      
\usepackage{inputenc}    
\usepackage{graphicx}    
\usepackage{booktabs}    
\usepackage{tabularx}
\usepackage[hypertex]{hyperref} 

\title{\textbf{WiP: Deception-in-Depth Using Multiple Layers of Deception}}
\author[1]{Jason Landsborough}
\author[2]{Neil Rowe}
\author[2]{Thuy Nguyen}
\author[1]{Sunny Fugate}

\affil[1]{NIWC Pacific, San Diego, CA, USA}
\affil[2]{Naval Postgraduate School, Monterey, CA, USA}

\date{April 2024} 

\begin{document}

\maketitle

\begin{abstract}
Deception is being increasingly explored as a cyberdefense strategy to protect operational systems.  We are studying implementation of deception-in-depth 
strategies with initially three logical layers: network, host, and data.
We draw ideas from military deception, network orchestration, 
software deception, file deception, fake 
honeypots, and moving-target defenses.  We are building a prototype representing our ideas and will be testing it in several adversarial  environments. We hope to show that deploying a broad range of deception techniques can be more effective in protecting systems than deploying single techniques.  Unlike traditional deception methods 
that try to encourage active engagement from attackers to collect intelligence, we focus on deceptions that can be used on real 
machines to discourage attacks.
\end{abstract}


\noindent\textbf{Keywords:} multilayer defense, deception-in-depth, cyber deception, honeypot

\maketitle

\section{Introduction}
Cyberattacks are increasing in quantity and sophistication, and increasingly affect our daily lives.  Cybersecurity was a 
\$153.65 billion market in 2022 \cite{fortune2023market}.  Despite all the current security 
products and decades of research, attackers are still getting into systems.  New ideas are needed.

Military organizations have experience with non-cyber attacks over many years, and may provide useful cyberdefense ideas.  
Nation-states are increasingly mounting cyberattacks against 
other nation-states and associated organizations.  Cyberattacks
can often be a force multiplier for traditional military operations.  
For example, in Operation GLOWING SYMPHONY, US Cyber Command targeted ISIS cyber assets to 
hurt their operations \cite{npr2019cybercom}.
Russia has also been using cyberattacks to target Ukraine, such as the CrashOverride malware that affected the power grid \cite{cisa2021crash}.  
The US military is often a target for cyberattacks. For example, 
Iran compromised a U.S. Navy Network which took defenders four months to repair by  
eliminating the attacker's access points \cite{iranNavyHack}.
Cyberattacks are also cost-effective for intelligence gathering, because organizations do not need a spy on the inside to steal information.  

Defensive deception 
is a promising way to aid cyber defenders.  
Deception can influence attackers to have false beliefs that impede their operations 
\cite{rowe2016introduction}, reducing their asymmetric advantage over defenders of surprise 
\cite{ferguson2019game, shade2020moonraker}.  Deception can also waste 
the attacker's time, a valuable resource in the civilian world as well as the military \cite{ferguson2018tularosa, ferguson-walter2021examining, michael2001intelligent}. 
These delays can give defenders more freedom of maneuver \cite{ferguson2019game}.  

Defensive cyber deception designed to encourage interaction often uses lures such as honeypots, 
but these are often easy to detect and ignore.
A more varied set of deceptions inspired by 
traditional cybersecurity and military concepts of defense-in-depth 
could provide a more balanced approach to deception, 
especially if it protects real systems.  In contrast to deceptions for encouragement 
that lure an attacker, such as traditional honeypots, discouraging deception tries to help real systems deter attacks.

\section{Cyber Deception Approaches}
\label{sec:related-work}
We describe here the cyber deception methods we think most effective in providing defense-in-depth.  Many cyber deception approaches come from military operations.  
Network orchestration technologies such as software-defined networking offer
opportunities to introduce deception into a network itself, which is useful
against the attacker doing network reconnaissance.  Host-based deception can be used 
after an attacker successfully exploits or accesses a system.  Data-based 
deception interferes with reconnaissance, lateral movement, or exfiltration and is often used with 
data objects 
such as fake files (honeyfiles).  Honeypots and honeytokens 
can often be easy to recognize, but there are ways to make it more difficult.  Two-sided 
deception is an approach to not only make honeypots look more realistic, but make 
real systems look fake.  Finally, moving-target defenses can work well with deception to randomize the 
attacker experience, causing delays, forcing repeated work,
and creating anxiety about possible detection.

\subsection{Military Deception}
Deception has been used by militaries for a long time.  The U. S. military uses a systematic approach 
to deception 
called \textit{MILDEC}, defined as  
``actions executed to deliberately mislead adversary military, 
paramilitary, or violent extremist organization decision makers, 
thereby causing the adversary to take specific actions (or inactions) that 
will contribute to the accomplishment of the friendly mission'' \cite{jp2012mildec}.
MILDEC plans to deliver deception techniques to a deception target through 
mechanisms called conduits to affect the target's quality of information.  MILDEC categorizes
techniques as either physical (such as the movement or training of forces),   
technical (using conduits such as media or computer networks), or administrative 
(using resources to signal or deny the deception).  

Military deception methods vary in how well they translate to cyberspace.  An assessment of techniques 
found that insights, lies, and concealment of intentions were the most appropriate for defense 
\cite{Rowe2004TwoTO, rowe2016introduction}.  Insights, or expectation of attacker actions, 
exploit how an attacker thinks and use that 
against them, such as knowing what information an attacker is likely to seek.  Lies, 
false responses to questions or requests, are useful 
because computer systems are usually very accurate in displaying facts true about 
a system to a user. Concealment of intentions
is done by honeypots to encourage interaction by an attacker.

\subsection{Deception with Network Orchestration}  
\label{sec:SDN}
Network orchestration is often done with 
software-defined networking. 
It uses two network planes: 
the data plane through which the traffic is routed and a control plane 
which modifies network devices. 
When used for deception, software-defined
networking can redirect traffic from a real network to a deceptive 
network \cite{shimanaka2019cyber}. 
It can rewrite packets and 
modify network flows to do this, such as with the OpenFlow software-defined 
networking framework.

The Chaos system used software-defined networking to 
assign random IP addresses to hosts, insert fake responses, and redirect traffic 
to a decoy server \cite{shi2017chaos}.  The deception strategy assessed the risk 
of an asset, and was more likely to do this to risky assets.
Risk can be judged by intrusion-detection systems, unexpected 
connections, or known vulnerabilities.

\subsection{Host-based Deception} 
Host-based deception focuses on an attack target.
Examples are intelligent software decoys that 
deceive only when an interface is being abused \cite{michael2001intelligent} or when an attacker or malicious 
code violates certain preconditions.  An attacker could 
see information that incorrectly suggests that they successfully exploited 
a piece of software.  Existing software can be made deceptive without modifying it by implementing a ``wrapper'' or outer routine around it.

A small-scale human-subject 
experiment tested the effectiveness of a Windows host-based deception 
 \cite{shade2020moonraker}.  The goal was to 
determine if the deception 
frustrated or slowed the attackers.  
The participants had information-technology
experience, and were given a network of workstations running simulated
supervisory control and data acquisition software 
as well as decoy hosts, and were asked to 
explore the network.   
The control group, interacting with real hosts, did perform better than the group that 
had deception present, indicating that the host-based deception was effective.  

\subsection{Data-based Deception}
\label{sec:data-based_honeytoken}
Honeytokens are objects to entice an attacker to reveal their capabilities or goals.  For example, a honeytoken can be a 
file showing a false network topology, which can entice attackers to use it in searching the network, receiving many error messages.
One company offers honeyfiles in a variety of formats which 
contain ``beacons'' that transmit data to the defender about attacker activities 
\cite{canarytokens}.  
Weaknesses of honeytokens are that beacons may only be
be triggered upon read of a file, and may not be triggered at all if read offline. Yuill et al. developed a honeyfile file system 
to better monitor access of these files \cite{yuill2004honeyfiles}.

Honeytokens for tabular data were studied by Shabtai et al.  \cite{shabtai2016behavioral}.  
One experiment created honeytokens containing
fake profiles of people with attributes such as age, country, eye color, 
job title, and education level; these could help track phishing and identity theft. 
Another experiment had students playing the role of 
a banker approving loans.  Bankers got a commission from loans they 
approved for the bank, and a larger commission if they misused the data by 
referring 
the customer to a private funding source.  Fake loan
requests that could earn large commissions were created using their honeytoken generator. 
They found that including the fakes in 20\% of the list of loans resulted in detecting 
100\% of the loans that
were referred to a private funding source, so the fakes helped detect illegal actions.  

\subsection{Fingerprinting Honeypots and Honeytokens}
\label{sec:fingerprinting}
Much research has examined ``fingerprinting'' 
(identifying honeypots and honeytokens) by
default configurations, their 
instrumentation or alerting mechanisms, or their missing features.  The Shodan security tool scans networks and reports information about Internet-connected devices; it has 
a proprietary algorithm to recognize
honeypots, which they call a Honeyscore \cite{shodan-honeyscore}.  
It provides a probability 
between 0.0 and 1.0 that a honeypot is at a given IP address.  A study of 
ICS honeypots found that Shodan's Honeyscore had a precision of 
70.3\%, using a Honeyscore of 0.5 or above as a positive result \cite{rowe2020effective}.

Another study tried to recognize instances of a
honeypot called GasPot that mimics a device 
which measures the level of a tank \cite{zamiri2019gas}.  It identified these  
characteristics of the device as relevant:
\begin{itemize}
	\item Use of the default configuration with well-known hard-coded 
	values.
	\item Implementation of only a subset of
	a protocol claimed by the device.
	\item Replies to queries with static values or highly predictable changes.
	\item Use of a traditional computing operating system instead of
	a more special-purpose operating system.
\end{itemize}  

With the above characteristics, the study could identify
17 GasPot instances on the Internet, whereas Shodan's Honeyscore 
 detected only 9, with no false 
 positives.  Gaspot responded quicker than real Automatic Tank Gauging devices
 which could also be used as an indicator.  
GasPot's output also indicated that it was running on 
Linux, since real devices used the carriage return and line feed for a new 
lines.  Gaspot originally only implemented 5 commands defined in the protocol used by real 
devices, which was also suspicious.

Thinkst offers a honeytoken service which 
they call Canarytokens.  They offer their honeytokens in formats such as Microsoft Office documents, PDF files, Web bug reports, QR 
codes \cite{canarytokens}.
A study tested the PDF  
honeytoken offered by Thinkst \cite{srinivasa2020towards}.  
Two techniques were used to detect the honeytoken: They 
sniffed network traffic to find 
DNS requests to the \textit{canarytokens.net} domain associated with Thinkst, and they parsed the PDF file and found an obfuscated 
link pointing to the same \textit{canarytokens.net} domain.  
The same authors found other canarytokens
had fingerprintable features, such as that AWS keys generated by Thinkst 
had the substring ``AKIAYVP4CIPP'' \cite{msaad2022honeysweeper}.  Furthermore, executable honeytokens
were all signed by the 
same certificate, and documents were created from the 
same base template with 
the same universally unique identifier, the same file size, and old file creation dates.

\subsection{Two-Sided Deception and Fake Deception}
\label{sec:two-sided}
Attackers can use tools to detect clues of 
deceptions such as honeypots and avoid interacting with them \cite{shodan-honeyscore, zamiri2019gas}.  
Defenders can use this desire to avoid honeypots to their advantage.  
A sophisticated attacker in particular, like an advanced persistent threat group, is especially unlikely to target a 
system they suspect is a 
honeypot \cite{shade2020moonraker, rowe2004model, rowe2007defending, 
franco2021survey}. Defenders 
can deliberately provide false clues of honeypots to steer attackers away from
real systems and towards safer systems that can further waste their time. 
Honeypots often have default parameter values or implement only a subset of a 
protocol.

Two-sided deception is a strategy where defenders make honeypots look 
more realistic while making real assets look less realistic.
It can be modeled by game theory \cite{miah2020concealing}, as in the honeypot selection game of Pibil et al., 
an extensive-form model which is extended to allow selection or modification of features on either a real or 
fake machine \cite{pibil2012game}.  Their work is motivated by
the signaling game of Carroll and Grosu \cite{carroll2011game}.
Their model consists of two machines and two features which 
represent ``signals'' that can be revealed to the attacker.  
The defender can change one feature (configuration) with an
associated cost, and the attacker observes the state of the machine.  
They find a small advantage with deception when
the cost to modify the features are the same.

\subsection{Moving-Target Defense and Deception}
\label{sec:mtd}
Much like conventional defenses, an attacker with sufficient time can learn 
enough about a system to recognize fakes. A way to prevent this is to use 
moving-target defenses in which the system environment such as 
IP addresses frequently changes.  It works much like a 
dealer in a casino shuffling a deck of cards to defeat card counting. 
It makes obsolete what the attacker has learned and requires them repeating 
reconnaissance and testing.  Moving-target defenses have been applied to
many aspects of cyber defense including software diversity 
\cite{larsen2014sok,homescu2013librando, landsborough2015removing}, n-variant 
program execution \cite{voulimeneas2021dmvx}, and network topology shuffling 
\cite{hong2017optimal}.

Moving target defenses are not necessarily deceptive, but deception can improve them by
obscuring information about what they are doing.    
The Chaos system \cite{shi2017chaos} described in section \ref{sec:SDN} is an 
example.  It assigned random IP addresses to risky hosts.  MT6D \cite{dunlop2011mt6d} used 
Internet Protocol address shuffling in the large address space of the IPv6 
protocol to conceal assets.  It routinely reassigned addresses, even 
with active connections in place. 

\section{Evaluating Cyber Defenses}
\label{sec:evaluating}
With so many deception options, it is important to evaluate their effectiveness.  Many methods can evaluate cyber defenses.  An expensive option is 
to use professional red teams to explore and try to compromise a system or network.  This
approach was taken in the Tularosa study on the effectiveness of deception \cite{ferguson2018tularosa}.  
This is not feasible for all organizations, but
tools can reduce the cost.  

``Breach and attack'' simulations can be used
to test defenses less expensively.  
Some are commercial products such as Cobalt Strike, and some are open-source.  
MITRE offers the Caldera tool (https://caldera.mitre.org/)
that manages actions with a centralized Command and Control server and uses a 
plugin modules to accomplish subtasks.  Another popular option 
is the Atomic Red Team that provides a library of tests  
for defenses (https://atomicredteam.io/).  
Both are open-source and mapped to the MITRE ATT\&CK framework. 

An organization called MITRE Enginuity offers attacking and defensive tool evaluations based on the 
ATT\&CK matrix.  In 2022 they 
evaluated two security vendors for deception, CounterCraft and SentinelOne 
\cite{mitre-evaluations}.
The attacker they modeled was APT29, which is affiliated with Russian intelligence.
The attacker code does not appear to use the Caldera framework but provides a collection 
of scripts.  
No evaluation score is given, but the released data shows the result of deception 
for various attacker techniques. 

MITRE offers two other useful tools for planning
active defenses including deception, each tied to
the ATT\&CK attack framework.  The D3FEND
matrix lists traditional passive defenses, 
and the ENGAGE matrix lists more active defenses.  These provide useful tactics 
for testing both passive and active defenses against various attacks.

\section{Research Methodology}
\label{sec:approach}
To research how new and existing deceptions can contribute to deception-in-depth, 
we will examine multilayered deception architectures. It is important to assess not only the effectiveness of the deceptions themselves, but how the success or failure of each deception affects the success or failure of the others. 

We identify three generic logical layers for deception: the network, the host, and 
the data. These represent areas where an attacker can 
interact with a system and acquire information.
We consider the network layer to be what is 
accessible externally over a network connection without user or kernel-space access.  
For example, routing of packets and querying a system with a network protocol would be 
network-layer activities.  Host-layer activities would typically involve a 
user interface such as a shell or graphical display on a host. 

Other deceptive layers can include deception on machines that act as a 
relay or dashboard for information, such as interfaces for control systems. 
Deception can be built into hardware as a physical layer of deception.  
We can also consider an attacker layer, where a defender ``hacks back'' the 
attacker, compromising their tools to introduce false information.  

Knowing what deceptions to use requires knowing what techniques an attacker might use.  We explored this 
in our earlier work examining advanced persistent threat techniques \cite{landsborough2024retrospectively}.  We used the MITRE APT groups dataset, 
where we identified the two most common techniques for a given tactic and  
candidate deceptions \cite{mitreAPTgroups}.  
The most promising deceptions 
were network deception, fake users, fake system information, and fake file-system 
tools.  We explored three publicly described attack campaigns to assess if deceptions could have been useful in thwarting
those attacks.  We also discussed modeling the connections between layers 
using conditional probabilities.  Certain deceptions are more effective 
at different layers, but attributes of the attacker's reaction to the deception are also important to 
consider.  Measures of the attacker's engagement include patience, adaptability, 
suspiciousness, alertness, and skill.

We describe the two phases of our proposed research below.

\subsection{Phase 1 Work -- Ongoing} 
First we must create an attack scenario based on the APT techniques 
identified in our prior work \cite{landsborough2024retrospectively}. 
We then must validate implemented deceptions using open-source tools such as Caldera 
mentioned in Section \ref{sec:evaluating}.  Open-source attacker simulation 
tools can help experiments by automating the processes of many trials in a controlled environment.
They could also help sharing the evaluation results or the  artifacts such as scripts and configurations with other
research groups.

We will identify ways to fingerprint some
common honeypots, and figure ways to provide
those clues to attackers while still enabling a system to be used for non-honeypot purposes, expanding the ideas in Section
\ref{sec:two-sided}.  This could be achieved using deceptive network orchestration inspired 
by work described in Section \ref{sec:SDN}.  

An efficient way to do this is to identify the characteristics of well-known honeypots, and our preliminary selection considered five factors:
\begin{itemize}
\item The honeypot must be \textit{open-source}.  Proprietary honeypots tend to be ``black boxes'' hard to analyze.  
\item The honeypot is \textit{popular}.  This increases the odds that an attacker may be familiar with it and knows how to detect it.  It may also mean that the honeypot runs services 
people have found useful, such as those similar to real machines in their network.
\item The honeypot is \textit{maintained}.  That means it is likely to work with 
updated versions of software and operating systems.  
\item The honeypot should be \textit{Linux-based} for ease of development and applicability to 
servers.  If the honeypot wants to look like a Windows machine or embedded system, it may have to modify features of the 
operating system besides the deceptive services they run.  
\item The honeypot has been publicly tagged as a honeypot by existing \textit{fingerprinting tools or approaches}.  This will simplify deceptions that discourage the attacker.  
\end{itemize}

Table \ref{tab:honeypot-candidates} 
compares some candidate honeypots to the above criteria.

\begin{table}[h]
  \centering
  \caption{Honeypots}
    \resizebox{\textwidth}{!}{%

  \begin{tabular}{lccccc}
    \toprule
    \textbf{Honeypot} & \textbf{Open source} & \textbf{Popular} & \textbf{Maintained} & \textbf{Linux-based} & \textbf{Fingerprinting} \\
    \midrule
    Kippo & Y & Y & Y & Y & Tool \\
    Cowrie & Y & Y & Y & Y & Approach \\
    (Thinkst) Open Canary & Y & Y & Y & Y & N? \\
    Glastopf & Y & Not anymore? & N & Y & Approach \\
    Conpot & Y & Y & N? & Y & Approach \\
    Dionaea & Y & Y & N & Y & Y \\
    (SANS) DShield & Y & N? & Y & Y & Unknown \\
    MongoDB-HoneyProxy & Y & N? & Y & Y & Unknown\\
    GasPot & Y & N? & Y & Y & Approach \\
    Honeyd & Y & N & N & Y & Approach \\
    \bottomrule
  \end{tabular}
  }
  \label{tab:honeypot-candidates}
\end{table}

Deception can also be designed against post-exploitation techniques such as 
living-off-the-land attacks 
on the host-based deception layer.  These are  are when an attacker 
uses existing system tools in their attack rather than downloading their own tools.  False system 
information or file system tools are the most promising.  But we also plan to 
get feedback from professional 
red-teamers within our professional network and also deploy honeypots 
to find other promising deceptions.  Variety 
in deceptions is essential to their effectiveness.

\subsection{Phase 2 Work} 
Because good attackers combine many techniques during an attack 
campaign, deception-in-depth should increase the effectiveness 
of defensive cyber deception interfering with an attacker.  
We will develop an architecture and criteria 
for how, why, and where to use multiple layers of deception 
for a given threat model.
A good candidate deception should be compatible with many
known attacker techniques, and have well-defined 
preconditions and postconditions so that 
defenders or automation can match them to possible attacks. 
Tracking deceptions and 
triggering events is also important for  interoperability, and 
may be tied to a defender's goal.
For example, observable actions by an attacker could weaken 
a deception or could be a form of signaling.  

Deceptions interact in several ways.  An attacker may not notice or react to a single deception, but multiple deceptions may create a cumulative effect in which an attacker does respond once the deceptions exceed a threshold.  
People often wait to make real-time decisions until enough pieces fall into place \cite{klein2017sources}, and this could be modeled as a threshold of cumulative deception.  

\begin{figure}[ht]
  \centering
  \includegraphics[width=\linewidth]{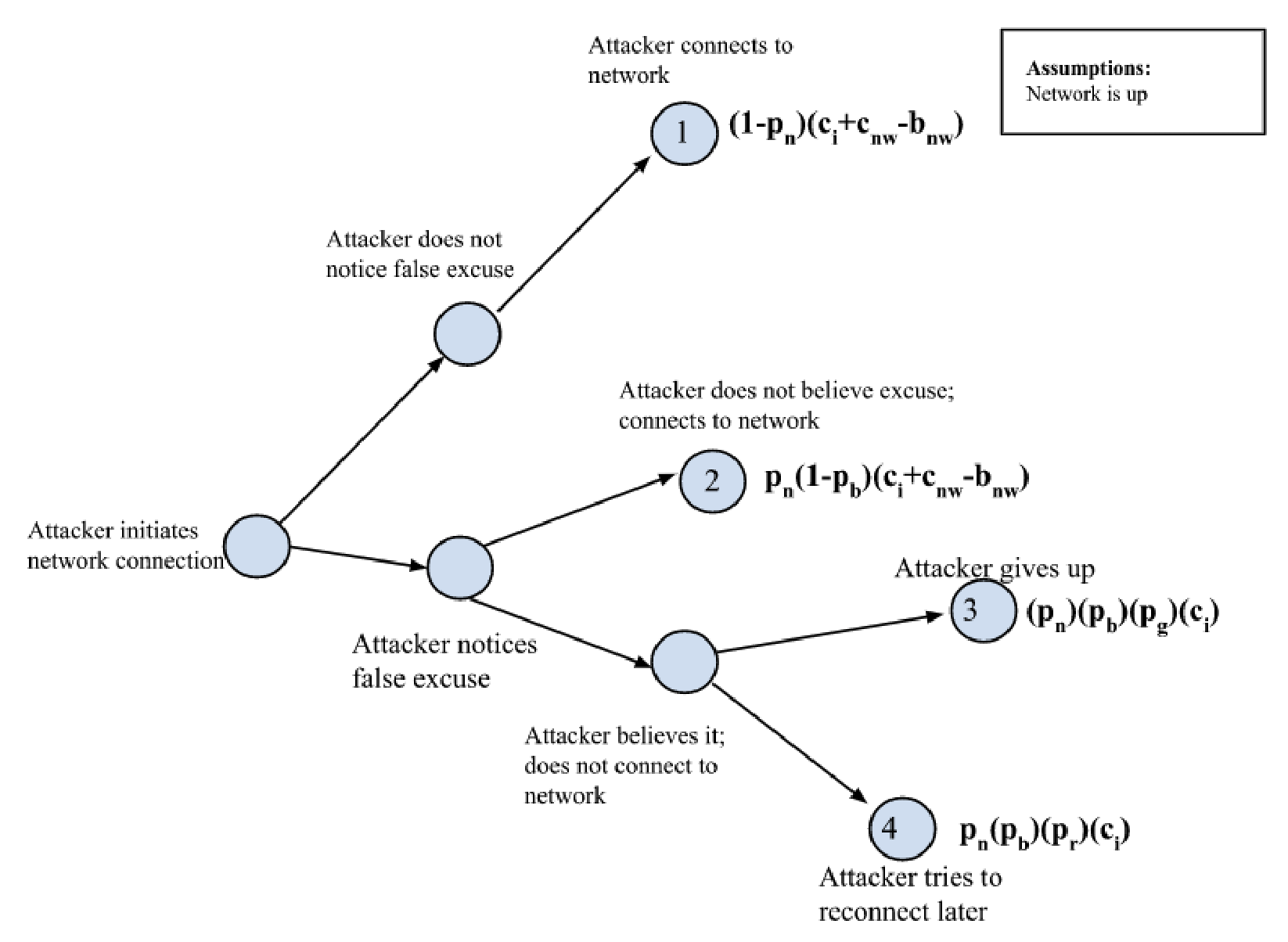}
  \caption{Decision tree for the ``network is down'' false excuse}
  \label{fig:NetworkIsDown}
\end{figure}

In general, we can assign conditional probabilities that deceptions work given that the attacker has seen another deception previously 
\cite{landsborough2024retrospectively}.  We can also assign costs to both the attacker and defender, and then reason about the resulting decision trees to find an optimal defense.
Figure \ref{fig:NetworkIsDown} shows a tree for a scenario where an attacker 
is presented with a false excuse that a network is down.  The costs to the attacker are 
c\textsubscript{i}, the cost of initial connection, and
c\textsubscript{nw}, the cost of maintaining the connection.  The benefit to 
the attacker is identified as b\textsubscript{nw}, the benefit of maintaining 
the network connection.
The probability of alertness 
is represented as p\textsubscript{n}, the probability the attacker notices
the false excuse.  The attacker's suspiciousness is represented as 
p\textsubscript{b}, the probability that the attacker believes the excuse.
Adaptability can be represented as p\textsubscript{g}, the probability that
the attacker gives up, and p\textsubscript{r}, the probability that the attacker
retries.  

Decision-tree analysis may be unnecessary for the many attackers that are looking
for easy targets, since simple deceptions may suffice in getting 
them to leave.  We will try to identify such deceptions to simplify planning.
Advanced persistent threat actors do not generally seek easy targets, but they
can still be risk-averse because surprising events interrupt their carefully
designed plans, and can cause them to waste resources.  
Some determined attackers enjoy a challenge and are unwilling to be 
discouraged from an attack.  They are similar to red-teamers who try many
attack methods to identify flaws in their target.  We need to learn clues to
identify the type of adversary that an attacker represents.

As systems become more complex, automation becomes as essential for defenders 
as for attackers.  Automated planning methods from artificial intelligence could
help and we will investigate them.
We still expect humans to have some supervisory control with such automation, 
however.  Coordinating tasks and boundaries will be important when multiple deception 
methods are combined, since that could lead to conflicting or sub-optimal 
solutions if deployed haphazardly.  ``Working agreements'' are a proposed technique allowing an operator and automated system to negotiate a suitable approach to sharing responsibilities.  This approach has promise for  defining role boundaries, improving calibration of trust, and improving transparency of decision-making
\cite{gutzwiller2018design}.  We believe this approach can be directly applied to cyber scenarios where well-defined 
roles and competencies can help guide multi-layer deception. 
Setting expectations is particularly true in a situation where the defender must successfully manage emergent properties of a complex composition of deceptions across many layers of the network and computing environment.  Working agreements seem useful 
for managing defenses in environments where there are impacts to legitimate users such as 
discouraging deceptions on a real server.  A technique like working agreements may also 
be useful for evaluating an automation's trust in defenders.  Fake or low quality working 
agreements can be presented to 
defenders to test their responses.  Enabling such working agreements could flag further scrutiny 
to determine if the defender is losing vigilance or is possibly an insider threat.
Zero-trust architectures now popular 
also may also offer new defensive opportunities because they can affect the relevance of deceptions.

\subsection{Evaluation}
We are planning three evaluation phases to validate our work.
In the first phase, we will validate our deceptions against standard attacker tools 
to determine if the tools convey the deception or crash.  
The second phase will be an adversarial assessment in an controlled environment.
This will likely use tools such as Caldera to automate experiments in an isolated 
lab environment.  This phase will also include control experiments where deception 
is not present.

The third phase will be an adversarial assessment in an uncontrolled environment. This may 
involve deploying deception to the Internet.  A challenge with deploying to the Internet 
is that the quality of the attacker will be unpredictable, and advanced persistent threats are unlikely to target a random host on the Internet.  
Internet-based evaluations can still offer results useful for defenders considering
deception in such an environment.

Evaluation metrics are important for measuring effectiveness of a defense plan.    
Whether the deception affected the attacker's behavior is one important measure.  
If deception is successful, it could cause the attacker to fail to achieve their important goals.  The type of failure state is also important to observe.  
With an accurate simulation of an attacker, we can measure timing characteristics of
their actions or the number of steps they
take, since often it is desirable to give defenders additional time to mitigate an attack by slowing the attackers rather than
stopping them abruptly.  

\section{Conclusion}
Defensive cyber deception is a promising tool to aid cyber defenders 
in protecting the security of their networks and systems.  
To increase chances of deception interfering with an attacker,
a more holistic deception-in-depth strategy is desirable.  
We discussed several deception methods that can be useful 
in a mutilayered deception approach.  We described our recent and ongoing 
work in examining attacker techniques, evaluation strategy, and popular honeypots.
We also discussed our research plan for developing a deception-in-depth architecture to aid defenders
in protecting real systems.
The
components and deployment considerations for deception-in-depth 
need to be studied more carefully to ensure such an approach can be used effectively.

\section*{Acknowledgements}
The views expressed are those of the authors and do not reflect the official policy or position of the Department of Defense or the U.S. Government.  This work was supported by Science, Mathematics And Research for Transformation (SMART) Scholarship for Service Program.  Angela Tan created the false excuse scenario and decision tree.

\bibliographystyle{plainurl} 
\bibliography{references} 

\begin{thebibliography}{10}

\bibitem{jp2012mildec}
Military deception.
\newblock Technical Report Joint Publication 3-13.4, 2012.

\bibitem{carroll2011game}
Thomas~E Carroll and Daniel Grosu.
\newblock A game theoretic investigation of deception in network security.
\newblock {\em Security and Communication Networks}, 4(10):1162--1172, 2011.

\bibitem{cisa2021crash}
CISA.
\newblock Crashoverride malware.
\newblock URL:
  \url{https://www.cisa.gov/news-events/alerts/2017/06/12/crashoverride-malware}.

\bibitem{dunlop2011mt6d}
Matthew Dunlop, Stephen Groat, William Urbanski, Randy Marchany, and Joseph
  Tront.
\newblock Mt6d: A moving target ipv6 defense.
\newblock In {\em 2011-MILCOM 2011 Military Communications Conference}, pages
  1321--1326. IEEE, 2011.

\bibitem{ferguson2019game}
Kimberly Ferguson-Walter, Sunny Fugate, Justin Mauger, and Maxine Major.
\newblock Game theory for adaptive defensive cyber deception.
\newblock In {\em Proceedings of the 6th Annual Symposium on Hot Topics in the
  Science of Security}, pages 1--8, 2019.

\bibitem{ferguson2018tularosa}
Kimberly Ferguson-Walter, Temmie Shade, Andrew Rogers, Michael
  Christopher~Stefan Trumbo, Kevin~S Nauer, Kristin~Marie Divis, Aaron Jones,
  Angela Combs, and Robert~G Abbott.
\newblock The {T}ularosa study: An experimental design and implementation to
  quantify the effectiveness of cyber deception.
\newblock Technical report, Sandia National Lab.(SNL-NM), Albuquerque, NM
  (United States), 2018.

\bibitem{ferguson-walter2021examining}
Kimberly~J. Ferguson-Walter, Maxine~M. Major, Chelsea~K. Johnson, and Daniel~H.
  Muhleman.
\newblock Examining the efficacy of decoy-based and psychological cyber
  deception.
\newblock In {\em 30th USENIX Security Symposium (USENIX Security 21)}, pages
  1127--1144. USENIX Association, August 2021.
\newblock URL:
  \url{https://www.usenix.org/conference/usenixsecurity21/presentation/ferguson-walter}.

\bibitem{franco2021survey}
Javier Franco, Ahmet Aris, Berk Canberk, and A~Selcuk Uluagac.
\newblock A survey of honeypots and honeynets for internet of things,
  industrial internet of things, and cyber-physical systems.
\newblock {\em IEEE Communications Surveys \& Tutorials}, 23(4):2351--2383,
  2021.

\bibitem{gutzwiller2018design}
Robert~S Gutzwiller, Sarah~H Espinosa, Caitlin Kenny, and Douglas~S Lange.
\newblock A design pattern for working agreements in human-autonomy teaming.
\newblock In {\em Advances in Human Factors in Simulation and Modeling:
  Proceedings of the AHFE 2017 International Conference on Human Factors in
  Simulation and Modeling, July 17--21, 2017, The Westin Bonaventure Hotel, Los
  Angeles, California, USA 8}, pages 12--24. Springer, 2018.

\bibitem{homescu2013librando}
Andrei Homescu, Stefan Brunthaler, Per Larsen, and Michael Franz.
\newblock Librando: transparent code randomization for just-in-time compilers.
\newblock In {\em Proceedings of the 2013 ACM SIGSAC conference on Computer \&
  Communications Security}, pages 993--1004, 2013.

\bibitem{hong2017optimal}
Jin~Bum Hong, Seunghyun Yoon, Hyuk Lim, and Dong~Seong Kim.
\newblock Optimal network reconfiguration for software defined networks using
  shuffle-based online mtd.
\newblock In {\em 2017 IEEE 36th Symposium on Reliable Distributed Systems
  (SRDS)}, pages 234--243. IEEE, 2017.

\bibitem{fortune2023market}
Fortune~Business Insights.
\newblock Cyber security market size, share \& covid-19 impact analysis, by
  security type (network security, cloud application security, end-point
  security, secure web gateway, application security, and others), by
  enterprise size (small \& medium enterprise and large enterprises), by
  industry (bfsi, it and telecommunications, retail, healthcare, government,
  manufacturing, travel and transportation, energy and utilities, and others),
  and region forecast, 2023-2030, 2023.
\newblock URL:
  \url{https://www.fortunebusinessinsights.com/industry-reports/cyber-security-market-101165}.

\bibitem{klein2017sources}
Gary~A. Klein.
\newblock {\em {Sources of Power: How People Make Decisions}}.
\newblock The MIT Press, 09 2017.
\newblock \href {https://doi.org/10.7551/mitpress/11307.001.0001}
  {\path{doi:10.7551/mitpress/11307.001.0001}}.

\bibitem{landsborough2015removing}
Jason Landsborough, Stephen Harding, and Sunny Fugate.
\newblock Removing the kitchen sink from software.
\newblock In {\em Proceedings of the Companion Publication of the 2015 Annual
  Conference on Genetic and Evolutionary Computation}, pages 833--838, 2015.

\bibitem{landsborough2024retrospectively}
Jason Landsborough, Thuy Nguyen, and Neil Rowe.
\newblock Retrospectively using multilayer deception in depth against advanced
  persistent threats.
\newblock In {\em HICSS}, 2024 (To appear).

\bibitem{larsen2014sok}
Per Larsen, Andrei Homescu, Stefan Brunthaler, and Michael Franz.
\newblock Sok: Automated software diversity.
\newblock In {\em 2014 IEEE Symposium on Security and Privacy}, pages 276--291.
  IEEE, 2014.

\bibitem{miah2020concealing}
Mohammad~Sujan Miah, Marcus Gutierrez, Oscar Veliz, Omkar Thakoor, and
  Christopher Kiekintveld.
\newblock Concealing cyber-decoys using two-sided feature deception games.
\newblock In {\em HICSS}, pages 1--10, 2020.

\bibitem{michael2001intelligent}
James~Bret Michael and Richard Riehle.
\newblock Intelligent software decoys.
\newblock {\em Engineering Automation for Reliable Software-Interim Progress
  Report (10/01/2000-9/30/2001)}, page~80, 2001.

\bibitem{mitreAPTgroups}
MITRE.
\newblock Groups.
\newblock URL: \url{https://attack.mitre.org/groups/}.

\bibitem{mitre-evaluations}
MITRE.
\newblock Apt29 deceptions, 2022.
\newblock URL:
  \url{https://attackevals.mitre-engenuity.org/trials-deceptions/apt29-deceptions/}.

\bibitem{msaad2022honeysweeper}
Mohamed Msaad, Shreyas Srinivasa, Mikkel~M Andersen, David~H Audran, Charity~U
  Orji, and Emmanouil Vasilomanolakis.
\newblock Honeysweeper: Towards stealthy honeytoken fingerprinting techniques.
\newblock In {\em Nordic Conference on Secure IT Systems}, pages 101--119.
  Springer, 2022.

\bibitem{npr2019cybercom}
NPR.
\newblock How the u.s. hacked isis, 2019.
\newblock URL:
  \url{https://www.npr.org/2019/09/26/763545811/how-the-u-s-hacked-isis}.

\bibitem{pibil2012game}
Radek P{\'\i}bil, Viliam Lis{\`y}, Christopher Kiekintveld, Branislav
  Bo{\v{s}}ansk{\`y}, and Michal P{\v{e}}chou{\v{c}}ek.
\newblock Game theoretic model of strategic honeypot selection in computer
  networks.
\newblock In {\em Decision and Game Theory for Security: Third International
  Conference, GameSec 2012, Budapest, Hungary, November 5-6, 2012. Proceedings
  3}, pages 201--220. Springer, 2012.

\bibitem{rowe2004model}
Neil~C Rowe.
\newblock A model of deception during cyber-attacks on information systems.
\newblock In {\em IEEE First Symposium onMulti-Agent Security and
  Survivability, 2004}, pages 21--30. IEEE, 2004.

\bibitem{rowe2007defending}
Neil~C Rowe, E~John Custy, and Binh~T Duong.
\newblock Defending cyberspace with fake honeypots.
\newblock {\em J. Comput.}, 2(2):25--36, 2007.

\bibitem{rowe2020effective}
Neil~C. Rowe, Thuy~D. Nguyen, Marian~M. Kendrick, Zaki~A. Rucker, Dahae Hyun,
  and Justin~C. Brown.
\newblock Creating effective industrial-control-system honeypots.
\newblock In {\em 53rd Hawaii International Conference on System Sciences},
  2020.

\bibitem{Rowe2004TwoTO}
Neil~C. Rowe and Hy~S. Rothstein.
\newblock Two taxonomies of deception for attacks on information systems.
\newblock 2004.
\newblock URL: \url{https://api.semanticscholar.org/CorpusID:142238729}.

\bibitem{rowe2016introduction}
Neil~C Rowe and Julian Rrushi.
\newblock {\em Introduction to cyberdeception}.
\newblock Springer, 2016.

\bibitem{shabtai2016behavioral}
Asaf Shabtai, Maya Bercovitch, Lior Rokach, Ya'akov Gal, Yuval Elovici, and
  Erez Shmueli.
\newblock Behavioral study of users when interacting with active honeytokens.
\newblock {\em ACM Transactions on Information and System Security (TISSEC)},
  18(3):1--21, 2016.

\bibitem{shade2020moonraker}
Temmie Shade, Andrew Rogers, Kimberly Ferguson-Walter, Sara~Beth Elsen, Daniel
  Fayette, and Kristin~E Heckman.
\newblock The moonraker study: An experimental evaluation of host-based
  deception.
\newblock In {\em HICSS}, pages 1--10, 2020.

\bibitem{shi2017chaos}
Yuan Shi, Huanguo Zhang, Juan Wang, Feng Xiao, Jianwei Huang, Daochen Zha,
  Hongxin Hu, Fei Yan, and Bo~Zhao.
\newblock Chaos: An sdn-based moving target defense system.
\newblock {\em Security and Communication Networks}, 2017, 2017.

\bibitem{shimanaka2019cyber}
Toru Shimanaka, Ryusuke Masuoka, and Brian Hay.
\newblock Cyber deception architecture: Covert attack reconnaissance using a
  safe sdn approach.
\newblock In {\em Proceedings of the 52nd Hawaii International Conference on
  System Sciences}, 2019.

\bibitem{shodan-honeyscore}
Shodan.
\newblock Honeypot or not?
\newblock URL: \url{https://honeyscore.shodan.io/}.

\bibitem{srinivasa2020towards}
Shreyas Srinivasa, Jens~Myrup Pedersen, and Emmanouil Vasilomanolakis.
\newblock Towards systematic honeytoken fingerprinting.
\newblock In {\em 13th International Conference on Security of Information and
  Networks}, pages 1--5, 2020.

\bibitem{canarytokens}
Thinkst.
\newblock Canary tokens, 2023.
\newblock URL: \url{https://canarytokens.org/}.

\bibitem{iranNavyHack}
The Verge.
\newblock Iranian hack of us navy network was more extensive and invasive than
  previously reported.
\newblock URL:
  \url{https://www.theverge.com/2014/2/18/5421636/us-navy-hack-by-iran-lasted-for-four-months-say-officials}.

\bibitem{voulimeneas2021dmvx}
Alexios Voulimeneas, Dokyung Song, Per Larsen, Michael Franz, and Stijn
  Volckaert.
\newblock dmvx: Secure and efficient multi-variant execution in a distributed
  setting.
\newblock In {\em Proceedings of the 14th European Workshop on Systems
  Security}, pages 41--47, 2021.

\bibitem{yuill2004honeyfiles}
Jim Yuill, Mike Zappe, Dorothy Denning, and Fred Feer.
\newblock Honeyfiles: deceptive files for intrusion detection.
\newblock In {\em Proceedings from the Fifth Annual IEEE SMC Information
  Assurance Workshop, 2004.}, pages 116--122. IEEE, 2004.

\bibitem{zamiri2019gas}
Mohammad-Reza Zamiri-Gourabi, Ali~Razmjoo Qalaei, and Babak~Amin Azad.
\newblock Gas what? {I} can see your gaspots. studying the fingerprintability
  of {ICS} honeypots in the wild.
\newblock In {\em Proceedings of the fifth annual industrial control system
  security ({ICSS}) workshop}, pages 30--37, 2019.

\end{thebibliography}

\end{document}